\documentclass[twocolumn,showpacs,prb,amsmath,amssymb]{revtex4}

\usepackage{graphicx}
\usepackage{epsfig}

\begin{document}
\title{Nuclear spin diffusion in quantum dots: Effects 
of inhomogeneous hyperfine interaction}
\author{Changxue Deng, Xuedong Hu}
\affiliation{Department of Physics, University at Buffalo, SUNY,
Buffalo, NY 14260-1500}
\date{\today}

\begin{abstract}
We study the effect of contact hyperfine interaction on the nuclear spin 
diffusion coefficients in semiconductor quantum dots.  The diffusion
coefficients are calculated with both the method of moment and density
matrix.  We show that nuclear spin diffusion is strongly suppressed by the
nonuniform hyperfine coupling resulting from the confined electron
wavefunction.  Our results agree with the observed suppression of nuclear
spin diffusion in these structures in recent experiments, and clarify the
degree of validity of the method of moment in an inhomogeneous system.   
\end{abstract}
\maketitle

\section{Introduction}

Nuclear spin polarization and dynamics \cite{Cory, Reimer} in semiconductor 
nanostructures such as
quantum wells and quantum dots have attracted increasing attention in recent
years.  For example, electrical transport experiments have demonstrated
dynamical nuclear spin polarization near tunnel junctions, quantum point
contacts, and coupled quantum dots.\cite{Kane,Wald,Huttel,Ono}  Optical
pumping nuclear magnetic resonance (NMR) technique has been used to explore
the local electronic state in 2D electron gas in the quantum Hall regime by
measuring the Knight shift and the relaxation time
$T_1$.\cite{science,exp-knight}  Nuclear spin diffusion has been found to
play an important role in the heat capacity anomaly\cite{exp-heat} at filling 
factor $\nu=1$, which may have originated from a Skyrme solid-liquid phase 
transition.  Time-resolved optical measurements in magnetic and non-magnetic 
semiconductor heterostructures also clearly demonstrate strong influences of
nuclear spins on the confined electron spin
dynamics.\cite{Gammon,Kawakami,Poggio}

Both nuclear\cite{QC-n1,QC-n2,Skinner} and electron spin\cite{QC-e} in
semiconductors have been proposed as the potential quantum bit for quantum
computing architectures, and nuclear spins also are suggested as quantum
memory.\cite{Taylor}  At low temperatures, the hyperfine interaction between
electron and nuclear spins could be the dominant decoherence mechanism for
both types of spins.\cite{dephase-dot,dephase-gas,Roger}  Because of the
confined nature of electrons in such devices, the hyperfine coupling acquires
a strongly local characteristics.  To achieve detailed understanding of
electron and nuclear spin coherence, a careful study of nuclear spin dynamics
in these semiconductor heterostructures is imperative.

One of the nuclear spin relaxation channels is spin diffusion, which reduces
local nuclear polarization through direct or mediated spin-spin interaction.
Nuclear spin diffusion (NSD) was first introduced by Bloembergen to explain
the measurements of spin-lattice relaxation time $T_1$ in ionic crystals in
the presence of paramagnetic impurities.\cite{Bloem} He suggested that NSD 
could be induced by the mutual nuclear spin flip-flops through magnetic
dipole-dipole interaction among nuclear spins.  Since then many
calculations\cite{many, Red} have been made for the NSD coefficients. 
Similar results were obtained via a variety of approaches, as these
calculations all deal with pure dipole-dipole interactions.

In this paper we present detailed calculations of NSD coefficients in
semiconductor quantum dots.  Although the formulation is general, we will
concentrate on GaAs based dots and wells which are of great experimental
interests.  Direct measurements of the NSD coefficients has been done using
optically pumped NMR for bulk GaAs and AlGaAs.  It was estimated that the NSD
coefficient in bulk GaAs is in the order of $10^{-13} cm^2/s$ for the arsenic
nuclei\cite{paget} and $\sim 10^{-14} cm^2/s$ for nuclei in the 
AlGaAs barrier.\cite{exp-well} 
Our objective in the present study is not to accurately predict the numerical
values of the NSD coefficients in the nanostructures.  Instead, we would like
to assess how they are modified compared to the bulk materials. 
Specifically, our present focus is on how the hyperfine interactions affect
the diffusion coefficients, since the confined electrons in these materials
have nonuniform wavefunctions, which lead to nonuniform coupling to the
nuclear spins through the Fermi contact interaction.  Since the hyperfine
interaction is much stronger than nuclear dipole-dipole interaction wherever
the electron wavefunction is not negligible, we expect that nuclear spin
diffusion could be strongly affected.

The paper is organized as follows.  In section II we introduce the moment
method\cite{Red} and the density matrix method,\cite{Lowe, ernst} which
we use to calculate the nuclear spin diffusion coefficients.  We then discuss
how to adapt these methods to the inhomogeneous situations of quantum dots. 
In section III, we give numerical results from both methods and compare the
two approximations.  We also explore the experimental relevance of our
results.  Finally some further discussion and conclusion are presented in
section IV.

\section{Formulation}
\subsection{Moment method}

In our calculation, we assume a finite static magnetic field $B_{0}$ along the
$z$ direction.\cite{parallel_field}  Under this condition non-secular terms
of dipolar Hamiltonian can be dropped due to consideration of energy
conservation, so that the direct magnetic dipolar Hamiltonian can be written
as\cite{Book}
\begin{eqnarray}
H_I & = & -\gamma_I \hbar B_0 \sum_i I_{iz} +\sum_{i\ne j} 
B_{ij}(2I_{iz}I_{jz} - I_{i+}I_{j-}), 
\label{eq:H-nuclear} \\
B_{ij} & = &
\frac{1}{4}\gamma_{I}^{2}\hbar^2R_{ij}^{-3}(1-3\text{cos}^2\theta_{ij}).
\label{eq:DD-coef}
\end{eqnarray}
Here $\gamma_I$ is the gyromagnetic ratio of nuclear spin $I$, $R_{ij}$ 
is the distance between two nuclei located at positions
$\mathbf{R_{\text{i}}}$ and $\mathbf{R_{\text{j}}}$, $\theta_{ij}$ 
is the angle between $\mathbf{R_{\text{ij}}}$ and $z$ direction, and  
$\sum_{i\ne j}$ stands for the summation over all the spin pairs except
$i=j$.  We will only consider the dipolar coupling among the same nuclear
species.  Effects of different spin species will be briefly discussed in
section IV.

The moment method was designed to study linear response of the spin
system\cite{Red}, such as the susceptibility of the nuclear spin system, by
applying a small space- and time-dependent magnetic field:
$$
b(x,t) = B_{1}\text{cos}(\omega t)\text{sin}(q x)\,,
$$
so that the response of the spin system can be evaluated.  To study spin
diffusion, the perturbing field is along the same direction as the static
magnetic field.  The perturbing Hamiltonian then takes the form
\begin{equation}
H_{1} = -\gamma_I \hbar B_{1}\text{cos}(\omega t)\sum_{i}
\text{sin}(qx_i)I_{iz}.
\label{eq:pert-H}
\end{equation}
The $2n$th moment is defined as\cite{Red,Book}
\begin{equation}
M_{2n} = \frac{\sum_{a,b}(E_{a}-E_{b})^{2n}
\vert\langle a \vert H_1|b\rangle\vert^2}
{\hbar^{2n}\sum_{a,b}\vert\langle a|H_1|b\rangle\vert^2},
\label{eq:2n-moment}
\end{equation}
where $a$ and $b$ are the eigenstates of the unperturbed nuclear spin
Hamiltonian and $E_a$ and $E_b$ are the associated eigenvalues. The moments
contain information on the shape of the resonance absorption curve for the
whole ensemble of nuclear spins.  A common practice is to assume a particular
line shape with some unknown parameters, then calculate the first few moments
to determine these parameters.\cite{Book}  In general, the calculation of
$M_{2n}$ is rather complicated.  However, knowing the first two moments is
usually enough to determine the line shape approximately.  In the present
situation, after substituting Eq.~(\ref{eq:pert-H}) into
Eq.~(\ref{eq:2n-moment}), we obtain
\begin{eqnarray}
M_2^{\mu\mu} & = & \frac{q^2}{2\hbar^2}\frac{\sum_{i\ne j}
x_{ij}^{\mu}x_{ij}^{\mu}\text{Tr}\{[H,I_{iz}][H,I_{jz}]\}}
{\sum_{i}\text{Tr}\{I_{iz}^2\}},
\label{eq:M2} \\
M_4^{\mu\mu} & = & -\frac{q^2}{2\hbar^4}
       \frac{\sum_{i\ne j} x_{ij}^{\mu}x_{ij}^{\mu}
\text{Tr}\{[H,[H,I_{iz}]][H,[H,I_{jz}]]\}}
            {\sum_{i}\text{Tr}\{I_{iz}^2\}},
\label{eq:M4}
\end{eqnarray}
where $\text{Tr}$ represents the thermal average of the operators, $x_{ij}^{\mu}=
x_i^{\mu}-x_j^{\mu}$ is the difference of the Cartesian coordinates at nuclear
sites $\mathbf{R}_i$ and $\mathbf{R}_j$, and Greek letters stand for the $x,
y$ and $z$ directions.  In deriving Eq.~(\ref{eq:M2}) and Eq.~(\ref{eq:M4}),
it is assumed that the nuclear spins are macroscopically homogeneous so that
$\sum_i I_{iz}$ commutes with the total Hamiltonian.  

NSD coefficients can be calculated starting from the general spin diffusion
equation,
\begin{equation}
\frac{\partial M(\bf{r},t)}{\partial t} = \sum_{\mu,\nu}D^{\mu\nu}
\frac{\partial ^2 M(\bf{r},t)}{\partial x^{\mu}\partial x^{\nu}}
\label{eq:spin-diff}.
\end{equation}
The diffusion of nuclear magnetization occurs as a result of a spatially
inhomogeneous initial condition of the magnetization.  As we mentioned above
the physical mechanism of NSD is the nuclear spin flip-flops.  For a known
line shape, we can calculate all the moments and evaluate the spin diffusion
coefficients.  In most cases the line shape can be approximated with a
Gaussian.  Using the Fourier transformed diffusion equation $\tau^{-1}=Dq^2$,
where $\tau$ is the polarization relaxation time;
the spin-diffusion coefficient $D$ can be expressed in terms of $M_2$ and
$M_4$\cite{Red}
\begin{equation}
D_{G}^{\mu\mu} = \frac{\sqrt{\pi}}{2}\frac{M_2^{\mu\mu}}{q^2} 
\left( \frac{M_2^{\mu\mu}}{M_4^{\mu\mu}} \right)^{\frac{1}{2}}.
\label{eq:diff-G}
\end{equation}
If $M_4^{\mu\mu}/3(M_2^{\mu\mu})^2$ is much greater than 1 (corresponding
to a long tail for the absorption line shape), the Gaussian approximation
becomes inappropriate.  A truncated Lorentzian shape with a large cutoff
frequency is usually assumed in such a situation.  The spin diffusion
coefficient $D$ is now
\begin{equation}
D_{L}^{\mu\mu} = \frac{\pi}{2\sqrt{3}} \frac{M_2^{\mu\mu}}{q^2} \left( 
\frac{M_2^{\mu\mu}}{M_4^{\mu\mu}} \right)^{\frac{1}{2}}.
\label{eq:diff-L}
\end{equation}
Since both $M_2$ and $M_4$ are proportional to $q^2$, the diffusion 
coefficients in expressions (\ref{eq:diff-G}) and (\ref{eq:diff-L}) 
are independent of $q$.  Notice that the two approximations of line shape lead
to almost the same numerical results for nuclear spin diffusion coefficients,
thus we adopt the Gaussian line shape (Eq.~\ref{eq:diff-G}) throughout this 
study.

In the present study we apply the moment method to study nuclear spin
diffusion in a quantum dot where trapped electrons are confined in all three
dimensions.  For simplicity we assume that there is only one electron in the
dot.  The nuclei-electron hyperfine interaction is given by
\begin{eqnarray}
H_h & = & \sum_i \rm{A}(\mathbf{R}_{i}) \, \mathbf{I}_{i} \cdot \mathbf{S},
\label{eq:H-HF} \\
\rm{A}(\mathbf{R}_{i}) & = & \frac{16\pi}{3}\gamma_I \gamma_e\hbar^2|
\Psi(\mathbf{R}_{i})|^2.
\label{eq:HF-coef}
\end{eqnarray}
Here $\gamma_e$ is the gyromagnetic ratio of the electron in the dot, and
$\Psi$ is the electron wavefunction.  In Eq.~(\ref{eq:H-HF}) we have ignored
the nuclei-electron dipolar interaction, which is much weaker than the
contact hyperfine interaction for any finite electron-nucleus distance.

The nuclear Zeeman energy splitting is about 0.2 percent of the electron
Zeeman energy in GaAs quantum dots.  Furthermore, the electrons in a quantum
dot has discrete energy spectrum.  There is no small change of electron
kinetic energy that could facilitate spin-dependent scattering.  Thus direct
spin flip-flops between the electron and nuclei are largely suppressed in
strong magnetic fields due to violation of energy conservation.  Here we also
neglect any phonon effect since it involves a higher order process and is 
not essential in the low temperature limit.  The hyperfine interaction in
Eq.~(\ref{eq:H-HF}) can now be reduced to the following effective Hamiltonian
(assuming electron spin is fully polarized.  A reduced electron spin
polarization will uniformly reduce the strength of $H_h$):
\begin{equation}
H_h = \frac{1}{2} \sum_i A_i I_{iz},
\label{eq:H-HFz}
\end{equation} 
where and $A_i=A(\mathbf{R_i})$, and the total Hamiltonian of the nuclear spin
system is
\begin{equation}
H_{\text{M}} = H_I + H_h,
\label{eq:H-M}
\end{equation}
where $H_I$ is the nuclear spin Hamiltonian given in Eq.~(\ref{eq:H-nuclear}). 
We notice here that similar approximation of neglecting electron-nucleus spin
flip-flop has also been used to calculate the electron spin spectral
diffusion induced by nuclear spin flip-flops.\cite{Roger}  In
Eq.~(\ref{eq:H-HFz}) we have ignored the spin dynamics of electron, and
assumed that the electron has been fully polarized.  Even if the average
electron polarization is zero, the calculation of the fourth moment in
Eq.~(\ref{eq:M4}) would still be non-vanishing, since the trace in
Eq.~(\ref{eq:M4}) involves terms like $\langle S_z^{2}\rangle=1/4$ and
$\langle S_z^{4} \rangle=1/16$. 

The calculation of moments has to be modified in the case of a quantum dot.
In a homogeneous nuclear spin system, the sum over nuclear spin site index $i$
in Eqs.~(\ref{eq:M2}) and (\ref{eq:M4}) is trivial because it means
calculating the average over the whole homogeneous sample.  For the
inhomogeneous system considered in the current study, we approximate the sum
over $i$ with the method of coarse graining where the sum is evaluated
over a few neighboring lattice points.  Such coarse graining is justified
since the strength of magnetic dipolar interaction decreases quite rapidly
($1/r^3$).

The calculation of the moments is greatly simplified at the high temperature
limit $k_BT \gg \hbar\gamma_I B_0$, which applies in most low temperature
experiments ($\sim 100$ mK electron temperature), since the nuclear Zeeman
energy is at the order of 1 mK/Tesla.  At the high temperature limit we can
neglect the Boltzmann factor in the thermal averages.  The actual evaluation
of the commutators and traces is long but straightforward.  The final results
are:
\begin{eqnarray}
\frac{\text{Tr}\{[H_{\text{M}},I_{iz}][H_{\text{M}},I_{jz}]\}}{\text{Tr}\{I_{iz}^2\}}
& = & \frac{4}{3} B_{ij}^2 I(I+1),
\label{eq:M2-QD} \\
\frac{\text{Tr}\{[H_{\text{M}},[H_{\text{M}},I_{iz}]][H_{\text{M}},[H_{\text{M}},I_{jz}]]\}}
{\text{Tr}\{I_{iz}^2\}} 
& = & M_{DD} + M_h, 
\label{eq:M4-QD}
\end{eqnarray}
where
\begin{eqnarray}
M_{DD} & = & \sum_{k(i,j)} \left\{ 3B_{ik}^2 B_{jk}^2 - 4B_{ij}^2[B_{ik}^2 +
B_{jk}^2 + (B_{ik}-B_{jk})^2] \right. \nonumber \\
& & + \left. 4B_{ij} B_{ik} B_{jk} (2B_{ij}-B_{ik}-B_{jk}) \right\}
\frac{32}{9} I^2 (I+1)^2 \nonumber \\
& & - \frac{8}{5}I(I+1)(16I^2+16I-7)B_{ij}^4, 
\label{eq:M4-QD-DD} \\
M_h & = & -\frac{2}{3}I(I+1) \, B_{ij}^2 \, (A_i-A_j)^2.
\label{eq:M4-Mh}
\end{eqnarray}
Here $\sum_{k(i,j)}$ means summation of $k$ over all the lattice points except
$i$ and $j$.  $M_{DD}$ and $M_h$ are the dipole-dipole contribution and
hyperfine contribution to the fourth moment, respectively.  Our results agree
with Redfield and Wu's results\cite{Red} if we set $A$ to be zero.

\subsection{Density matrix method}

Since the moment method is designed for study of homogeneous bulk system, it
is important to corroborate our results on inhomogeneous systems with a
different approximation.  As a comparison, we calculate diffusion
coefficients using the density matrix method,\cite{ernst, Lowe} which is more
straightforward in terms of its physical picture.  We assume that the density
matrix of nuclear spin system can be expanded in terms of a group of
orthogonal operators $I_i~(i=1,N)$, where $N$ is the number of nuclear spins
in the system: 
\begin{equation}
\rho = \sum_i a_i(t) I_{iz},
\label{eq:approx_DM}
\end{equation}
and
\begin{equation}
\text{Tr}\{I_{iz}I_{jz}\} = \delta_{ij}\text{Tr}\{I_{iz}^2\}.
\end{equation}
This choice for the nuclear spin density matrix is a good approximation at the
high temperature limit, which is usually satisfied by the systems we are
interested in.  The total Hamiltonian for the nuclear spin system is
\begin{eqnarray}
&&H_{\text{DM}} = H_0 + H_1, \nonumber \\
&&H_0 = \sum_i (\frac{1}{2}A_i-\gamma_I\hbar B_0)I_{iz} 
+ 2\sum_{i\ne j} B_{ij}I_{iz}I_{jz}, \nonumber \\
&&H_1 = -\sum_{i\ne j}B_{ij}I_{i+}I_{j-}.
\end{eqnarray}
Here we take the nuclear flip-flop term (which accounts for spin diffusion) as
a perturbation. 

The nuclear spin density matrix can be conveniently calculated in the
interaction picture 
\begin{equation}
\tilde{\rho}(t) = e^{iH_0t}\rho(t)e^{-iH_0t}.
\end{equation}
We have used ``$~\tilde{}~$'' to represent the operators in the interaction
picture.  The temporal dynamics of the density matrix in the interaction
picture is governed by the flip-flop term in the full Hamiltonian
\begin{equation}
\dot{\tilde{\rho}}(t) = -i[\tilde{H}_1(t),\tilde{\rho}(t)].
\end{equation}
A second-order calculation leads to
\begin{equation}\label{rho}
\dot{\tilde{\rho}}(t) = i[\tilde{\rho}(t), \tilde{H}_1(t)] + i^2 
\int_0^{t} d\tau[\tilde{H}_1(t),[\tilde{H}_1(t-\tau),\tilde{\rho}(t)]].
\end{equation}
Substituting Eq.~(\ref{eq:approx_DM}) into Eq.~(\ref{rho}), we find
\begin{equation}
\dot{a}_k(t) = \sum_i W_{ki}a_i(t),
\end{equation}
with
\begin{eqnarray}
W_{ki} &=& \frac{1}{\text{Tr}[I_{kz}^2]}
\left\{ -i \text{Tr}\left[[\tilde{H}_1(t),I_{iz}]I_{kz} \right] \right.
\nonumber  \\
& + & \left. \int_0^t d\tau \text{Tr} \left[ [\tilde{H}_1(t),I_{kz}]
[\tilde{H}_1(t-\tau),I_{iz}] \right] \right\}.
\label{wik}
\end{eqnarray}
One can easily show that $W_{ki}=W_{ik}$ by noting that the trace is invariant
under the cyclic reordering of operators.  $W_{ik}$ describes the flip-flop
rate of two nuclear spins at site $i$ and $k$.  Substituting the density
matrix $\rho(t) = \sum_i^Na_i(t)I_{iz}$ into the equation of motion of the
local nuclear magnetization
\begin{eqnarray}\label{eq:eom}
\frac{\partial}{\partial t}\langle I_{kz} \rangle
 &=& \text{Tr}\left\{ \dot{\rho}(t) I_{kz}\right\}
 = \text{Tr}\left\{ \dot{\tilde{\rho}}(t) I_{kz}\right\} \nonumber \\
 &=& \sum_i W_{ki}\langle I_{iz} \rangle,
\end{eqnarray}
and performing Taylor expansion around the space point of the $k\text{th}$
nucleus, we find
\begin{equation}\label{eq:eom1}
D_{\mu\nu}^k = \frac{1}{2}\sum_{i(k)}W_{ik}(x_i^{\mu} - x_k^{\mu})(x_i^{\nu} -
x_k^{\nu}).
\end{equation}
In writing Eq.~(\ref{eq:eom}) we have used the commutation relation $[I_{kz},
H_0]=0$.  It is easy to show that $W_{kk}\approx 0$, because it involves a
summation of a fast oscillatory function that averages to zero over many
nuclear sites.  Physically, $W_{kk}$ corresponds to energetically impossible
processes and has no physical meaning.  It then follows that the zero-order
term in the Taylor expansion does not contribute to spin diffusion.  The
first-order term also vanishes because of the crystal
symmetry.\cite{note_cubic} 

To calculate the diffusion coefficients we need to find the flip-flop rates
$W_{ik}$.  The explicit calculations of the traces for an arbitrary nuclear
spin in Eq.~(\ref{wik}) are quite complicated.  In the following we consider
the particular situation of spin 3/2 nuclei, which is the case for GaAs
quantum dot.  Calculating the trace for $I=3/2$, we obtain
\begin{eqnarray}
&&\text{Tr} \left\{ [\tilde{H}_1(t), I_{kz}] [\tilde{H}_1(t-\tau), I_{iz}]
\right \} = 2B_{ik}^2\text{cos}(\frac{A_{ik}}{2}\tau) \nonumber \\
&&f(4B_{ik}\tau)
\prod_{m(i,k)} 2\left[ \cos(2B_{ikm}\tau) + \cos(6B_{ikm}\tau) \right],
\end{eqnarray}
where $f(x)=34+48\text{cos}(x)+18\text{cos}(2x)$. Here we have used the
definition $A_{ik} = A_i - A_k$ and $B_{ikm} = B_{im} - B_{km}$. Finally we
get the expression of $W_{ik}$
\begin{eqnarray}\label{eq:Wik}
W_{ik} &=& \frac{B_{ik}^2}{10} \int_0^t d\tau \cos(\frac{A_{ik}}{2}\tau)
f(4B_{ik}\tau) \nonumber \\
& & \times \prod_{m(i,k)} \cos(4B_{ikm}\tau) \cos(2B_{ikm}\tau)\,,
\end{eqnarray}
which would then allow us to calculate the NSD coefficient of the system.

\section{Numerical results}

\subsection{NSD in bulk system}

Before presenting our numerical results for a quantum dot, we first estimate 
the NSD coefficients for pure nuclear spin dipole-dipole interaction using
Eq.~(\ref{eq:H-nuclear}) with the moment method.  Notice that the hyperfine
interaction does not change the first moment.  The summations in
Eq.~(\ref{eq:M4-QD-DD}) can be easily done over the nuclei in a face-centered
cubic structure (for GaAs).  Since the dipole interaction decays as $r^{-3}$,
the summations converge quite rapidly.  A numerical calculation yields
$D^{zz} = 0.29 \gamma_I^2 \hbar/a_{\text{GaAs}}$ and $D^{xx} = 0.16
\gamma_I^2 \hbar/a_{\text{GaAs}}$ for $I=\frac{3}{2}$, where the lattice
constant $a_{GaAs}=5.65 $~\AA.  These values are comparable to Lowe and
Gade's results\cite{many} for spin one half in a simple cubic structure.  For
the specific example of $^{75}$As nuclei, where $\gamma_I = 4.58 \times 10^3
\frac{1}{\text{s} \cdot \text{G}}$, $D^{zz} = 1.1 \times 10^{-13}
\text{cm}^2/s$ and $D^{xx} = 6.3 \times 10^{-14} \text{cm}^2/s$.  Spin
diffusion is faster along the $z$ direction because the dipolar interaction
is stronger along the external magnetic field direction according to
Eq.~(\ref{eq:DD-coef}).  Specifically, the dipolar coupling coefficient is
proportional to the magnitude of $1-3\text{cos}^{2}\theta_{ij}$.  Along $z$
direction this value is -2, while it is 1 along $x$ or $y$ direction.  In the
following discussion we use $D_0^{\mu\mu}$ to represent the NSD coefficient
for pure dipole-dipole interaction in the absence of inhomogeneity.  

To calculate the NSD coefficients with the density matrix method, we have to
evaluate the integral in Eq.~(\ref{eq:Wik}).  This can be done by first
changing the upper limit of the integration to infinity because the integrand
is a product of many cosine functions that has a sharp spectral peak near
$\tau=0$, so that changing the integration upper limit only introduces 
a negligible error.\cite{Lowe}  We thus have
\begin{eqnarray}
& & \int_0^t du ~\prod_{i=1}^{N}\text{cos}(a_i u) \nonumber \\
& = & \int_0^t du ~\text{exp}\left( \text{ln}\prod_{i}\text{cos}(a_i u)
\right) \nonumber \\
& \approx & \int_0^{\infty} du~\text{exp}\left( -\frac{1}{2}au^2 \right)
= \frac{1}{2}\sqrt{\frac{2\pi}{a}},
\label{second_step}
\end{eqnarray}
where $a=\sum_i^N a_i^2$. In the second step of the calculation in
Eq.~(\ref{second_step}) we have expanded the integrand around $u=0$ and kept
only the terms to the order $\text{O}(u^2)$.  This approximation is in the
same spirit as the steepest descent method. Using this approximation we find 
Eq. \ref{eq:Wik} takes the following form
\begin{eqnarray}
W_{ik} &=& F_{ik}^{(0)} + F_{ik}^{(1)} + F_{ik}^{(2)}, \nonumber \\
F_{ik}^{(0)} &=& \frac{17\sqrt{2\pi}}{5}B_{ik}^2
(A_{ik}^2 + g_{ik} )^{-\frac{1}{2}}, \nonumber \\
F_{ik}^{(1)} &=& \frac{12\sqrt{2\pi}}{5}B_{ik}^2
( A_{ik}^2 + 64B_{ik}^2 + g_{ik} )^{-\frac{1}{2}}, \nonumber \\
F_{ik}^{(2)} &=& \frac{9\sqrt{2\pi}}{10}B_{ik}^2
(A_{ik}^2 + 256B_{ik}^2 + g_{ik} )^{-\frac{1}{2}},
\nonumber \\
g_{ik} &=& 80\sum_{p(i,k)}(B_{ip}-B_{kp})^2. 
\end{eqnarray}
The calculated $W_{ik}$ can then
be inserted into Eq.~(\ref{eq:eom1}) to obtain the diffusion coefficients. 
For pure dipolar interaction we find $D^{zz} = 0.49 \gamma_I^2
\hbar/a_{\text{GaAs}}$ and $D^{xx} = 0.21 \gamma_I^2 \hbar/a_{\text{GaAs}}$. 
These calculated $D^{zz}$ and $D^{xx}$ are nearly twice as large as the
results given by the moment calculations, although we do find that $D^{zz}$
is greater than $D^{xx}$, similar to the results of the moment calculations.

\subsection{NSD in a quantum dot}

We now include electron-nuclear spin hyperfine interaction in our calculation
of the NSD coefficients.  To study the effects of hyperfine interaction, we
need knowledge of the electronic wavefunctions.  The ground state electron
wavefunction in a 2D gated GaAs quantum dot can be approximated by
\begin{eqnarray}
\Psi(\mathbf{r}) & = & \frac{u({\bf r})}{\sqrt{\pi}l_0}\sqrt{\frac{2}{z_0}}
\text{cos} \left( \frac{\pi z}{z_0} \right) \, e^{-\frac{1}{2l_0^2}(x^2+y^2)},
\label{eq:Psi} \\
l_0 & = & l_B r_0 (l_b^4+r_0^4/4)^{-\frac{1}{4}},\nonumber 
\end{eqnarray}
where $z_0$ is the quantum dot thickness, $l_0$ is the Fock-Darwin radius, and
$r_0$ is the electrostatic lateral parabolic confinement radius.  The value of
the $\Gamma$-point Bloch function $u({\bf r})$ at nuclear sites can be deduced
from experimental measurements.\cite{exp-U}

\begin{figure}[t]
\begin{center}
\epsfig{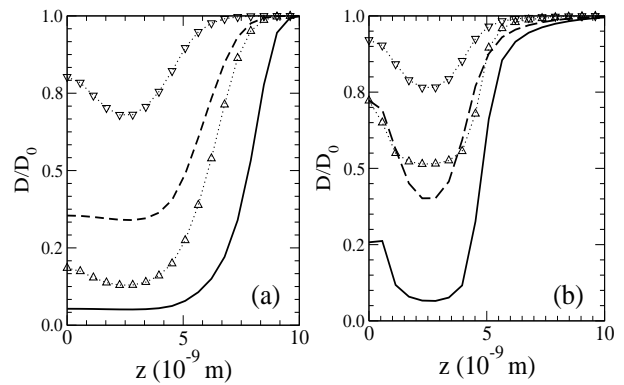} 
\caption{
\label{fig1}
The ratio of NSD coefficients $D^{\mu\mu}/D_0^{\mu\mu}$ as a function of
spatial coordinate $z$ for various Fock-Darwin radii, where $z$ is the
perpendicular distance from the center of the dot.  The left panel (a) shows
the results using moment method while the right panel (b) represents those
obtained with density matrix method. In all these calculations we assume a
quantum dot with thickness $z_{0}$ = 10 nm.  The solid line (upward triangle)
represents $l_0=30$ nm for $D^{zz}$ ($D^{xx}$) and the dashed line (downward
triangle) describes $l_0=80$ nm for $D^{zz}$ ($D^{xx}$).  }
\end{center}
\end{figure}

\begin{figure}[t]
\begin{center}
\epsfig{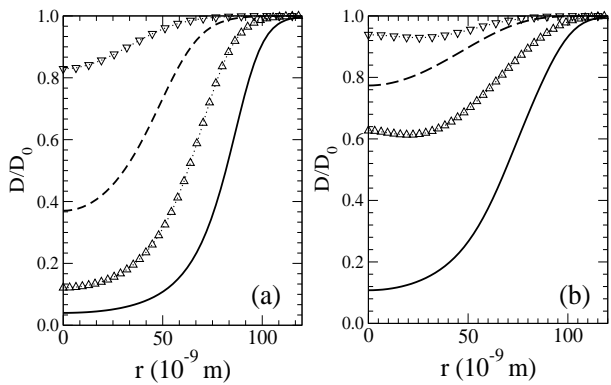} 
\caption{
\label{fig2}
The ratio of NSD coefficients $D^{\mu\mu}/D_0^{\mu\mu}$ as a function of
spatial coordinate $r$ for various quantum dot thickness, where $r$ is the
radial displacement in the 2D plane.  The origin of the coordinate system is
located at the center of the dots. The left panel (a) are the results of 
moment method and the right panel (b) shows those from density matrix method.
In all these  calculations we have used a quantum dot with Fock-Darwin radius 
$l_{0}$ = 50 nm. The solid line (upward triangle) describes $z_0=5$ nm 
for $D^{zz}$ ($D^{xx}$) while the dashed line (downward triangle) shows
 $z_0=15$ nm for $D^{zz}$ ($D^{xx}$).
} 
\end{center}
\end{figure}

\begin{figure}[b]
\begin{center}
\epsfig{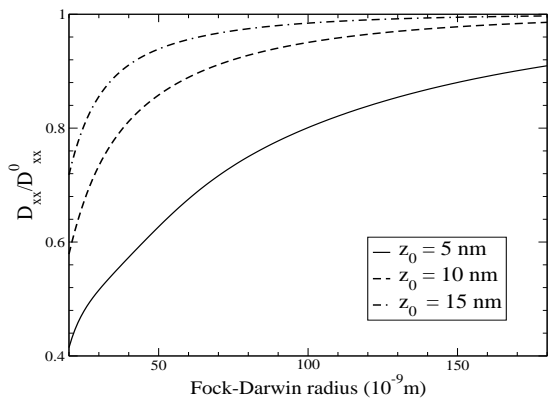} 
\caption{
\label{fig3}
The ratio of NSD coefficients $D^{xx}/D_0^{xx}$ at the center of the dot as 
a function of Fock-Darwin radius $l_0$ with three different dot thickness. The
results are obtained with the density matrix method.} 
\end{center}
\end{figure}

The calculated NSD coefficients using the moment method and density matrix
approach share several common characteristics.  Figure~\ref{fig1} shows both
$D^{zz}/D_0^{zz}$ and $D^{xx}/D_0^{xx}$ as functions of spatial coordinate
(along the external magnetic field) $z$ for two different quantum dot
thickness.  In Fig.~\ref{fig2} we plot $D^{zz}/D_0^{zz}$ and
$D^{xx}/D_0^{xx}$ as functions of the radial displacement $r$ (perpendicular
to the external magnetic direction) for different Fock-Darwin radius $l_0$. 
The curves in both figures show similar behaviors.  The suppression of spin
diffusion due to hyperfine interaction at or near the center of the quantum
dot could be so significant that $D^{zz}$ and $D^{xx}$ is only a few percent
of $D^{zz}_0$ and $D^{xx}_0$.  Figures~\ref{fig1} and \ref{fig2} also show
that with both methods the suppression of spin diffusion decreases as the dot
size becomes larger, which can be explained by noting that hyperfine
interaction strength decreases for larger dots.  To further illustrate this
point, in Fig.~\ref{fig3} we show the diffusion coefficient $D^{zz}$ at the
center of the quantum dot as a function of Fock-Darwin radius $l_0$.  Similar
results (which are not shown in Fig.~\ref{fig3}) are found for $D^{xx}$ as
well.  An additional feature of Fig.~\ref{fig1} is that, at the boundary of
the dot along $z$ direction, the NSD coefficients increase to $D_0$ rapidly. 
This behavior is due to our assumption that the electron wavefunction outside
the dot is zero.  It is also noticed that $D^{zz}$ has a stronger suppression
than $D^{xx}$, which is illustrated in both Fig.~\ref{fig1} and
Fig.~\ref{fig2} using both calculation methods.

There are several interesting differing features in the results of the two
methods in addition to the different magnitudes of the diffusion coefficients 
given by the two methods as we have discussed previously for homogeneous
systems.  Firstly, moment method typically leads to stronger suppressions at
the center of the dot.  Secondly, Fig.~\ref{fig1}, particularly panel (b),
indicates that the suppression of NSD is not the strongest at the center of
the quantum dot.  Instead it decreases from the center (albeit only slightly
in some cases), reaches its minimum at an intermediate position, then starts
to rise again near the edge of the dot.  This feature is stronger in the
results obtained through the density matrix method than those from the moment
method, and is stronger for the $z$-direction diffusion (Fig.~\ref{fig1}) than
in-plane directions (Fig.~\ref{fig2}).  In fact the characteristic exists,
although only weakly, in Fig.~\ref{fig2}(b) for $D^{xx}$ while it is not
present in Fig.~\ref{fig2}(a).  

The presence of off-center local minimum in nuclear spin diffusion in a
quantum dot is physical.  The suppression of spin diffusion is determined by
the difference of the inhomogeneous hyperfine coupling at two nuclear sites
$\propto |\Psi_i|^2 - |\Psi_j|^2 \sim \nabla |\Psi_i|^2 \cdot ({\bf r}_i -
{\bf r}_j)$ instead of the coupling constant alone (see Eq.~(\ref{eq:M4-Mh})
and Eq.~(\ref{eq:Wik})).  In essence, in the flip-flop processes that account
for nuclear spin diffusion, energy must be conserved.  If the hyperfine
coupling strengths are not the same at the two lattice sites, the effective
Zeeman energies of the two nuclear spins are different, so that extra energy
must be absorbed or emitted (from the overall dipolar energy reservoir, for
example) to compensate for the difference.  Apparently small energy
differences should result in greater flip-flop rates, hence faster diffusion. 
This is the basic reason why spin diffusion is suppressed in an inhomogeneous
system.  Since the energy difference is proportional to both the magnitude
and the gradient of the electron wavefunction, the strongest suppression of
NSD coefficient could occur either at the center of a quantum dot or its
``waist'', where the gradient is the largest (notice that the nuclear spins
are on discrete sites, thus near the center of the quantum dot the hyperfine
energy difference is generally nonvanishing, for a very small dot it could
even be a maximum depending on the form of the electron wavefunction).  For
an electron confined in a quantum dot geometry, the envelope wavefunction is
highly non-uniform and can be approximated with Eq.(~\ref{eq:Psi}).  Along
the $z$ direction the changes of this hyperfine coupling between neighboring
nuclear sites are quite large.  This is the reason that there is a sharp
minimum of the diffusion coefficient as a function of $z$ at a nonzero $z$. 
On the other hand the confinement is not that strong in the $r$ direction
especially for larger dots, so the feature is not as obvious.

\begin{figure}
\begin{center}
\epsfig{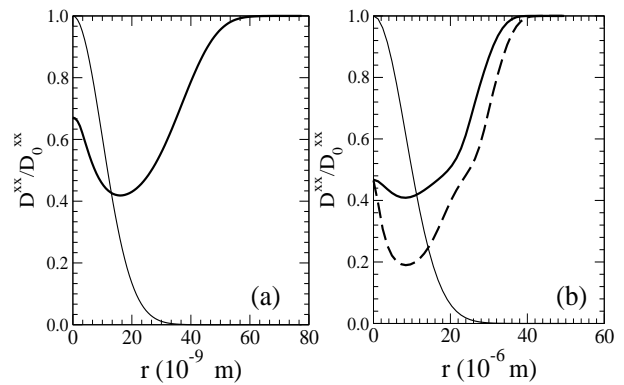} 
\caption{
\label{fig4}
The ratio of NSD coefficients $D^{xx}/D_0^{xx}$ (calculated using the density
matrix method) as a function of spatial coordinate $r$ for strongly confined
dots (panel a) and quantum wells (panel b) with Gaussian shaped carrier spin
polarization (see discussion in the text).  In the case of quantum dots we
assume $z_0=10$ nm and $l_0$ = 25 nm.  In the calculation of quantum well
with thickness of $15$ nm, we assume the effective radius of the Gaussian
shaped carrier spin polarization is 12 $\mu$m. In both panels the thinner
lines show the variations of hyperfine coupling strength as a function of
$r$.  In panel b, the two solid and dashed lines represent different
polarization magnitude.
} 
\end{center}
\end{figure}

To illustrate the previous discussion in more detail we consider a small
quantum dot with $l_0 = 25$ nm and $z_0 = 10$ nm.  Figure~\ref{fig4}(a)
clearly shows a sharp off-center local minimum of the NSD coefficient
$D^{xx}$, corresponding to a strong suppression of diffusion, which is weak
in Fig.~\ref{fig2}.  Quite interestingly a ring structure of nuclear spin
polarization has been observed in ferromagnet/semiconductor heterostructures
by spatially modulating the excitation intensity.\cite{ring}  The experiment
uses a laser pulse with a Gaussian cross section, which we believe is key to
the ring structure.  The inhomogeneous power input induces non-uniform
carrier polarization strength at the interface which in turn leads to
inhomogeneous hyperfine couplings.  As we have discussed, the suppression of
diffusion would be the strongest at some position between the center and the
boundary.  In Fig.~\ref{fig4}(b), we show the suppressions of NSD coefficient
$D^{xx}$ for two different polarization magnitudes in the micrometer size. 
The results are quite similar to the strongly confined quantum dots.  At low
temperatures an important spin relaxation mechanism is spin diffusion.  The
existence of the local minimum of diffusion coefficients contributes to a
maximum of nuclear spin polarization since the diffusion is slowest at the
point.  In other words a ring structure could very well be present, as what
was observed experimentally.

\section{Discussion and conclusion}

In the present study we have investigated the dipole-dipole interaction
among like nuclear species (Eq. (\ref{eq:H-nuclear})).  Interaction between
unlike nuclear spins have been neglected.  Under the assumption that the
magnetic field is not weak, this should be a good approximation.  However
there is the so-called indirect interaction (RKKY)\cite{Book} in highly
disordered samples where spin-flip scattering has measurable physical
effects.  In this regard, the coupling between different nuclear species may
have non-negligible effects.  It should be mentioned that $M_2$ and $M_4$ in
Eq.~(\ref{eq:M2-QD}) and Eq.~(\ref{eq:M4-QD}) do not change without direct
spin interaction.  However, the evaluation of $M_4$ becomes extremely
complicated if the indirect coupling between unlike spin species is included. 
We did not study this aspect of spin diffusion in the current paper. 
 
We have calculated NSD coefficients for Arsenic nuclei in this paper.  GaAs
has a zinc blende structure with 50\% $^{75}$As (which is the only stable 
Arsenic isotope).  In natural GaAs samples, there are two isotopes of Gallium,
$^{71}$Ga (19.8\%, $\gamma_I = 8.16\times 10^3 1/\text{s} \cdot \text{G}$)
and $^{69}$Ga (30.2\%, $\gamma_I = 6.42 \times 10^3 1/\text{s} \cdot
\text{G}$).  In the barrier region, the Ga concentration is even lower with
the introduction of 10\% to 15\% of Al in place of Ga.  An evaluation of the
NSD coefficients for Ga would have to account for the random distribution of
different Ga isotopes on the fcc lattice.  Here our emphasis is the effect of
inhomogeneous hyperfine interaction on NSD.  Furthermore, this nonuniform
hyperfine coupling, in the form of Eq.~(\ref{eq:H-HFz}), cannot compensate
for the difference in Zeeman energy of different nuclear species, so that the
inter-species NSD is unlikely.  For example, the effective hyperfine magnetic
field seen by nuclei at the center of a quantum dot is only a few tens Gauss,
which is usually much less than the external field.  Thus the inter-species
NSD is basically impossible in a finite magnetic field, and we do not have to
consider the Ga nuclei when studying NSD of the As nuclei.

Recently spin diffusion suppression by nonuniform field has been
found\cite{MRFM} in silica samples where an inhomogeneous magnetic field was
generated by a ferromagnetic tip of a magnetic resonance force microscope. 
It was found that spin relaxation rates $T_{1}^{-1}$ are significantly
reduced due to the suppression of nuclear spin flip-flop processes.  In
solids with paramagnetic impurities,\cite{Bloem} inhomogeneous internal field
could also be generated by dipole-dipole interaction between the impurity and
its neighboring nuclear spins, in which case a barrier to NSD can also be
formed.

In this study we have presented two methods to study the suppression of spin 
diffusion.  In the moment method originally developed for homogeneous bulk 
material,\cite{Red} a small spatially and temporally varying perturbation is
added to the total Hamiltonian to generate linear response.  The spatial
variation of the perturbation is assumed to be smooth compared to electron 
wavefunction variation (long wavelength approximation). 
To apply it to nanostructures like quantum dots, we combine
it with a coarse-graining approximation. In this method we have to assume the 
sizes of quantum dots being considered are relatively large.
In principle the moment method is
not designed for the calculation of strong spatial variation of diffusion. 
This partly explains the less prominent local minimum of diffusion
coefficients as a function of spatial coordinates.  On the other hand the
density matrix method is more straightforward and keeps more local features
in the evaluations. With our density matrix calculation no assumption of any
particular line-shape is necessary. We simply start from the full Hamiltonian,
and then use the second-order perturbation to find the evolution of the local 
nuclear magnetization.    
Nevertheless the strong suppressions of NSD coefficients
for small dots appear in both calculations. 
 
To conclude we have presented detailed study of nuclear spin diffusion under
the influence of inhomogeneous contact hyperfine interactions in GaAs based
nanostructures.  Our results show that there are strong suppressions of spin
diffusion at the center and the waist of a quantum dot or quantum well where
electron probability and/or gradient of electron probability is large, which
is consistent with experimental observations in such
structures.\cite{exp-heat,ring}  The numerical results given in Section III
show that NSD coefficients could be suppressed to as small as a few percent of
$D_0^{\mu\mu}$.  Our results clearly show that non-uniform electron
distribution can help maintain desired nuclear spin polarization in these
semiconductor nanostructures.


\begin{thebibliography}{99}
\bibitem{parallel_field} In all our calculations the applied magnetic field is
parallel to the growth direction (vertical field) of the quantum dots or
quantum wells.  If the magnetic field is in an in-plane direction (parallel
field), as long as it does not lead to resonance between electron and nuclear
spins, we do not anticipate qualitative differences in terms of nuclear spin
diffusion.
\bibitem{note_cubic} In the case of GaAs quantum dot which is of the most
theoretical and experimental interests, the lattice has cubic symmetry.
Combining this with the equation $W_{ik}=W_{ki}$, one can quickly show that
the first-order term vanishes.  See also Ref.~\onlinecite{Lowe}.
\end{thebibliography}
\end{document}